\documentclass{article}

\usepackage{arxiv}

\usepackage[utf8]{inputenc} 
\usepackage[T1]{fontenc}    
\usepackage{hyperref}       
\usepackage{url}            
\usepackage{booktabs}       
\usepackage{amsfonts}       
\usepackage{nicefrac}       
\usepackage{microtype}      
\usepackage{lipsum}		
\usepackage{graphicx}
\usepackage{natbib}
\usepackage{doi}

\title{ARES OS 2.0: An Orchestration Software Suite for Autonomous Experimentation Systems and Self-Driving Labs}

\date{April 3, 2026}	

\author{ \href{https://orcid.org/0000-0002-7066-1678}{\includegraphics[scale=0.06]{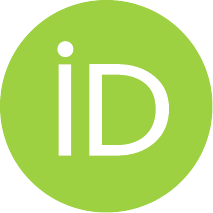}\hspace{1mm}Arthur     W. N. Sloan}\textsuperscript{1, 2}\\
    \And
    \href{https://orcid.org/0000-0001-6958-2932}{\includegraphics[scale=0.06]{orcid.pdf}\hspace{1mm}Robert W. Waelder}\textsuperscript{1, 3} \\
    \And
    \href{https://orcid.org/0009-0003-1849-7051}{\includegraphics[scale=0.06]{orcid.pdf}\hspace{1mm}Morgen L. Smith}\textsuperscript{1, 3, 4} \\
    \And
    Nicholas Kleiner\textsuperscript{5} \\
    \And
    Arnas Babeckis\textsuperscript{5} \\
    \And
    Jason Wheeler\textsuperscript{5} \\
    \And
    Daylond Hooper\textsuperscript{5} \\
    \And
    \href{https://orcid.org/0000-0002-7066-1678}{\includegraphics[scale=0.06]{orcid.pdf}\hspace{1mm}Benji 
    Maruyama}\textsuperscript{1,*}\\
}

\renewcommand{\shorttitle}{ARES OS 2.0}

\hypersetup{
pdftitle={A template for the arxiv style},
pdfsubject={q-bio.NC, q-bio.QM},
pdfauthor={David S.~Hippocampus, Elias D.~Striatum},
pdfkeywords={First keyword, Second keyword, More},
}

\begin{document}
\maketitle

\textbf{1} \textit{Air Force Research Laboratory, Materials \& Manufacturing Directorate, United States of America} \\
\textbf{2} \textit{The National Research Council, United States of America}\\
\textbf{3} \textit{AV Inc., United States of America}\\
\textbf{4} \textit{Kansas State University, Tim Taylor Department of Chemical Engineering, United States of America}\\
\textbf{5} \textit{DCS Corp., United States of America}\\


\section{Summary}
\label{sec:summary}
\verb|ARES OS 2.0| (hereinafter \verb|ARES OS|) is an open-source software suite to enable laboratory automation and closed-loop autonomous experimentation. Its function is to orchestrate experimental actions and data handoff between lab equipment, analysis routines, and experimental planning modules through a service-oriented architecture. \verb|ARES OS| is abstracted to apply to general experimental flows common in materials science, chemistry, and biology and related disciplines. The core of \verb|ARES OS| provides central control over all modules, along with the heavy lifting of UI creation, data management, and experimental design tools. \verb|ARES OS| modules communicate with the core software over\verb|protobuf| and \verb|gRPC|, allowing them to be language-agnostic and user-creatable. This allows users to easily implement modules that control experimental hardware, process collected data , or plan experiments to meet their specific research needs. \verb|ARES OS| lowers the barrier to entry for researchers to build their own self-driving labs, allowing them to focus on scientific programming for their use case and reducing the effort and time needed to bring an autonomous experimentation system online. 

\section{Statement of Need}
\label{sec:statement_of_need}

\begin{figure}
	\centering
    \includegraphics{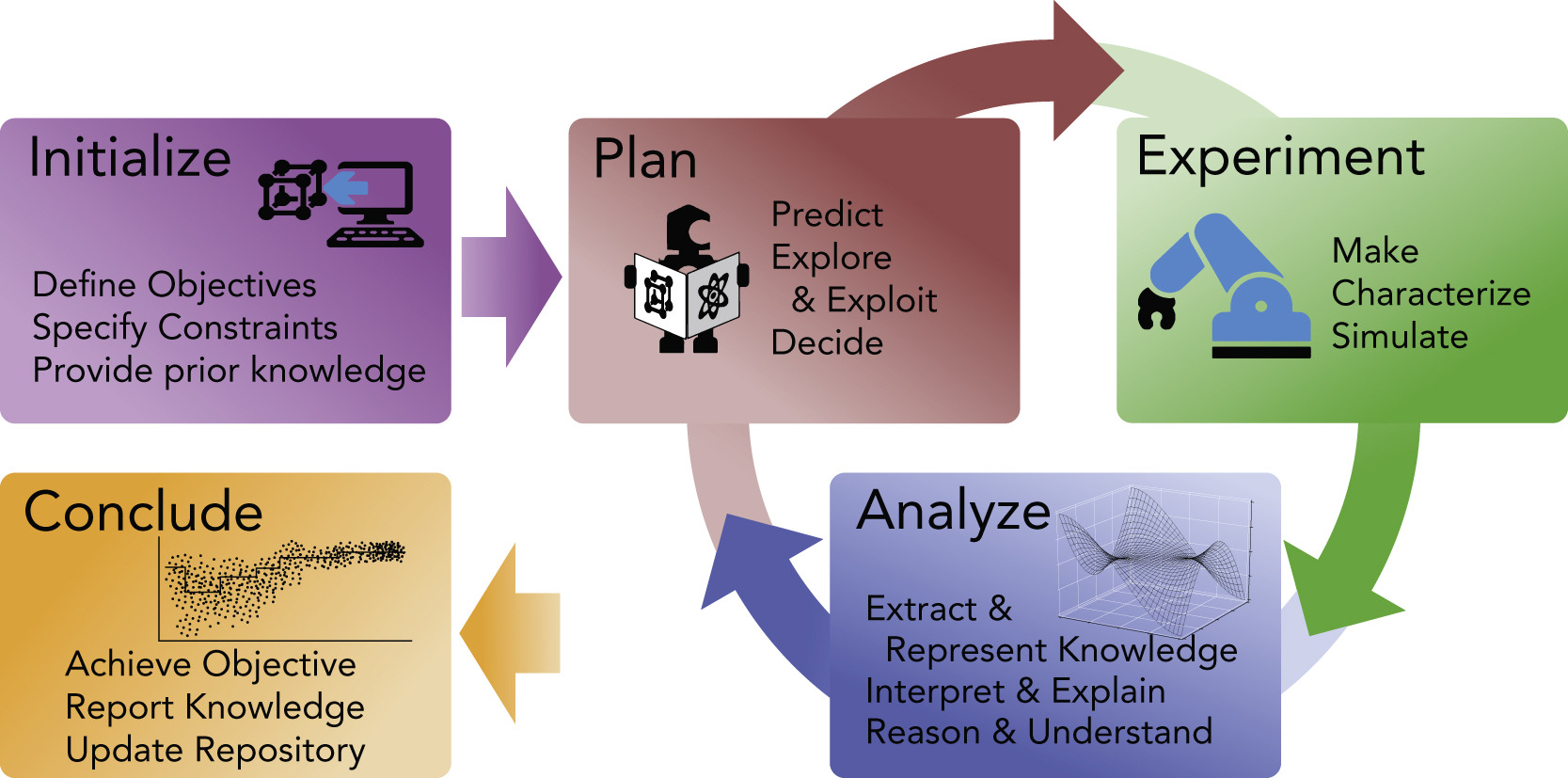}
	\caption{A closed loop, research autonomy process flow. Used with permission from \citep{stach:2021}. Copyright Elsevier 2021}
	\label{fig:fig_1}
\end{figure}

Research and technology development in the physical sciences has historically been a slow, expensive, and labor-intensive process. To overcome these issues and accelerate the pace of discovery, researchers across a variety of fields have started a revolution in how science is done: Autonomous Experimentation (AE). AE, also called self-driving labs (SDLs), combines robotic high throughput experimentation (HTE) techniques with in situ and in-line analysis methods, and artificial intelligence/machine learning (AI/ML) planning routines to autonomously plan, execute, and analyze experiments in pursuit of a user defined goal \citep{stach:2021, abolhasani:2023}, with the objecting of making scientific research faster, better, and cheaper. This process flow is shown in \autoref{fig:fig_1}. AE systems have been demonstrated to provide faster research progress, lower experimental variability, and a reduced number of experiments to reach a goal compared to traditional manual planning and experimentation \citep{stach:2021}. Our group published the first autonomous experimentation system, ARES, for materials in 2016, and \verb|ARES OS| has been in development since \citep{nikolaev:2016}.

Implementing a new SDL traditionally has a high barrier to entry, with software being a major contributor\citep{lo:2024}. Today there is a growing number of SDL orchestration software offerings, and while some are low cost and/or open-source, they tend to be specific to a research domain. Many SDLs rely either on expensive offerings from commercial vendors, or are bespoke, researcher-built systems, with long implementation timelines due to the complexity and array of technical disciplines required to successfully develop and integrate all aspects of an SDL \citep{lo:2024} (e.g., software architecture, mechatronics, AI/ML, domain specific scientific knowledge). These factors pose a high barrier to entry and constrain SDL development to well-funded research organizations, slowing the application of SDLs to new research problems. 

From a researcher standpoint, the largest hurdle in developing an SDL is the integration of separate elements into a functioning autonomous system \citep{seifrid:2022}. Thanks to the wealth of data analysis, ML, and other scientific libraries available, many researchers have sufficient competence with Python to create the individual modules of autonomous system but may lack the software engineering expertise to integrate them in a robust and flexible manner. `ARES OS` was developed to address this core issue by providing researchers with a modular framework for coordinating hardware, software, and data management. This framework is combined with an easy-to-use, self-populating UI and companion Python library, \verb|ARES OS| \citep{PyAres}, which allows users to rapidly develop, test, and integrate system components.

\section{State of the Field}
\label{sec:state_of_the_field}
Several other open-source SDL orchestration software packages are available. Notable examples include MadSci\citep{MADSci}, ChemOS2.0\citep{sim2024}, and Minerva-OS\citep{zaki2025}. Compared to these alternatives, \verb|ARES OS| differentiates itself primarily through the researcher-first user experience, which places an emphasis on low- or no-code operation of core features. All interactions with the core functionality of \verb|ARES OS| can be accomplished within the GUI. This includes installation, analyzer/planner module configuration, hardware control, building and executing experimental campaigns, and data export. \verb|ARES OS| has also been successfully abstracted to several different domains of scientific research including additive manufacturing, wet chemistry, and chemical vapor deposition, while other software offerings may to be more specialized to a single domain.

\section{Software Design}
\label{sec:software_design}
\verb|ARES OS| uses a service-oriented architecture with a C\# and ASP .NET core, written to follow SOLID principles for understandability, flexibility, and maintainability. The core application handles the backend logic necessary for automation and autonomy, such as experimental routines, database interactions (\verb|ARES OS| supports SQL Server, SQLite, and Postgres), and provides frameworks for interacting with system modules, such as custom GUIs, laboratory hardware, experimental planners, and data analyzers. 

Communication between the core and system module services is facilitated by Google's protobuf and gRPC. The use of protobuf allows for easy data transmission over the network, facilitating the use of both local and remote experimental or computing resources. Protobuf also allows ARES OS to be language-agnostic, enabling the creation or re-use of modules written in any supported language (e.g., C\#, Python, Javascript, R, etc.). 

By default, \verb|ARES OS| includes a Blazor UI, designed as an intuitive hub for customizing and using an AE system, allowing for both centralized computer control of experimental hardware and the execution of user-defined campaigns for automated or autonomous experimentation. The \verb|PyAres| library is available via PyPi and provides an easy-to-use interface to create and configure \verb|ARES OS| compatible devices, planners, and analyzers with only a few lines of Python code \citep{PyAres}. For ease of use we have also created an \verb|ARES OS| launcher application, which streamlines the installation and configuration of \verb|ARES OS| and the necessary databases and certificates \citep{ARES-Launcher}. The \verb|ARES OS| launcher also supports installation from specific forks of \verb|ARES OS| to enable users to develop modified versions that fit their specific use cases.

\section{Research Impact Statement}
\label{sec:research_impact_statement}
\verb|ARES OS| was designed primarily to be used by experimental researchers in the physical sciences for the implementation of AE/SDL systems. Additionally, \verb|ARES OS| is suitable for use by students for use in a classroom setting to study ML and AE principles. As part of its development, `ARES OS` has been used in experimental systems to study a variety of materials science problems such as carbon nanotube synthesis \citep{waelder:2024, bulmer:2023} and fused deposition modeling 3D printing \citep{deneault:2021}. ARES OS will also be used in new curriculum under development by the University of Buffalo’s department of Materials Design and Innovation.

\section{Availability}
\label{sec:availability}
The `ARES OS` Core source code is available from the public GitHub repository (\url{https://github.com/AFRL-ARES/ARES/releases}). The `ARES OS` launcher source code is available from the public GitHub repository (\url{https://github.com/AFRL-ARES/ARES-Launcher/releases}) with downloadable binaries for Linux, Windows and MacOS. The `PyAres` companion library is available on PyPi (\url{https://pypi.org/project/PyAres}) for installation with \verb|pip| or from the public GitHub repository (\url{https://github.com/AFRL-ARES/PyAres/releases}).

\section{Acknowledgments}
\label{sec:acknowledgments}
The authors gratefully recognize funding from the Air Force Office of Scientific Research under LRIR 25COR019, R. Doug Riecken, PO.

\bibliographystyle{unsrtnat}
\bibliography{references}  
\end{document}